\documentclass{blois}

\usepackage{amsmath,amssymb,amsthm,amsfonts,amsbsy,latexsym}
\usepackage{slashed}
\usepackage{tikz}
\usetikzlibrary{shapes.misc}
\usetikzlibrary{arrows,shapes}
\usetikzlibrary{trees}
\usetikzlibrary{decorations.pathmorphing}
\usetikzlibrary{decorations.markings}
\usetikzlibrary{decorations.pathreplacing}
\usetikzlibrary{calc}

\def\be{\begin{equation}}
\def\ee{\end{equation}}
\def\bea{\begin{eqnarray}}
\def\eea{\end{eqnarray}}

\newcommand{\Photo}{}
\newcommand{\tn}[1]{\textnormal{#1}}
\newcommand{\hhref}[1]{\href{http://arxiv.org/abs/#1}{arXiv:#1}}

\begin{document}
\tikzset{
  x=1pt, y=1pt,
  vector/.style={decorate, decoration={snake}, draw},%
  provector/.style={decorate, decoration={snake,amplitude=2.5pt}, draw},%
  antivector/.style={decorate, decoration={snake,amplitude=-2.5pt}, draw},%
  fermion/.style={postaction={decorate}, decoration={markings,mark=at position .55 with
      {\arrow[]{>}}}},%
  fermionbar/.style={draw=black, postaction={decorate}, decoration={markings,mark=at position .55
      with {\arrow[draw=black]{<}}}},%
  fermionnoarrow/.style={draw=black},%
  scalar/.style={dashed,draw=black, postaction={decorate}, decoration={markings,mark=at position .55
      with {\arrow[draw=black]{>}}}},%
  scalarbar/.style={dashed,draw=black, postaction={decorate}, decoration={markings,mark=at position
      .55 with {\arrow[draw=black]{<}}}},%
  scalarnoarrow/.style={dashed},%
  electron/.style={draw=black, postaction={decorate},
    decoration={markings,mark=at position .55 with {\arrow[draw=black]{>}}}},%
  bigvector/.style={decorate, decoration={snake,amplitude=4pt}, draw}, 
  gluon/.style={decorate, decoration={coil,amplitude=4pt, segment length=5pt}},%
  bigarrow/.style={draw=black, postaction={decorate}, decoration={markings,mark=at position 1 with
      {\arrow[draw=black]{>}}}},%
  stealth/.style={>=stealth, thick},%
  myshade/.style={inner sep=0pt, draw=gray, blur shadow={shadow blur steps=20, shadow blur extra rounding}},
  spy using mag glass/.style={spy scope={every spy on node/.style={circle,fill, fill opacity=0.0, text opacity=1},
      every spy in node/.style={magnifying glass, fill=white, draw=black, circular drop shadow,ultra thick, cap=round},#1}},%
}%
\hfill DESY 15-172
\vspace*{4cm}
\title{BOOSTED HIGGS CHANNELS}

\author{ MATTHIAS SCHLAFFER}

\address{DESY, Notkestrasse 85, D-22607 Hamburg, Germany}

\maketitle\abstracts{In gluon fusion both a modified top Yukawa and new colored particles can alter
  the cross section. However in a large set of composite Higgs models and in realistic areas of the
  MSSM parameter space, these two effects can conspire and hide new physics in a Standard Model-like
  inclusive cross section.\\
  We first show that it is possible to break this degeneracy in the couplings by demanding a boosted
  Higgs recoiling against a high-$p_T$ jet. Subsequently we propose an analysis based on this idea
  in the $H\to2\ell+E\!\!\!/_T$ channels. This measurement allows an alternative determination of
  the important top Yukawa besides the $t\bar t H$ channel.}

\section{Introduction}
\label{sec:introduction}

The top quark and its coupling to the Higgs play a central role in the hierarchy problem. Many
models for physics beyond the Standard Model (BSM) addressing this issue predict a modified top
Yukawa coupling and its precise measurement can thus give crucial input to the search for BSM
dynamics.\par
Two important processes for this measurement are $t\bar t h$ and gluon fusion. The former is
difficult to measure due to the high multiplicity final state while the latter has a sizable cross
section despite being loop suppressed. However gluon fusion is not only altered by a modified top
Yukawa coupling but also by new colored particles, e.g.~scalar tops or composite top partners, that
can run in the loop besides the ordinary top quark.\par
If the new loop particles are heavier than the Higgs, $m_h^2/4 \,m_{loop}^2 \ll 1$, their
contribution to the gluon fusion process can be described by an effective gluon-gluon-Higgs
interaction\cite{HLET}
\begin{equation}
  \label{eq:1}
 \mathcal{L}_{eff}=\kappa_g\frac{\alpha_S}{12\pi v} G_{\mu\nu}^a G^{a\,\mu\nu}h\,,
\end{equation}
where $\kappa_g$ is a coefficient quantifying the size of the interaction with $\kappa_g=0$
corresponding to the Standard Model (SM), $\alpha_S$ is the strong coupling constant,
$v\approx 246\,\textnormal{GeV}$ the Higgs vacuum expectation value, and $G_{\mu\nu}^a$ the gluon
field strength tensor. The modified top Yukawa can be easily accounted for by multiplying the top
Yukawa term by a new coefficient $\kappa_t$, where $\kappa_t=1$ corresponds to the SM.  Yet the mass
relation $m_h^2/4\,m_t^2\ll 1$ is fulfilled for the top quark and thus the top induced gluon fusion
process can be described alternatively by Eq.~(\ref{eq:1}) with $\kappa_g$ replaced by
$\kappa_t$. Consequently the inclusive gluon fusion cross section is given by
$\sigma_\tn{incl}(\kappa_t,\kappa_g)/\sigma_\tn{incl}^{\tn{SM}}\approx \left(\kappa_t+\kappa_g\right)^2$
with corrections to this formula being beyond the reach of the LHC\cite{Gillioz:2012se}.\par
Therefore the inclusive gluon fusion process at the LHC cannot be used to disentangle the two
coefficients. An independent measurement of $\kappa_t$ and $\kappa_g$ is however important as new
physics could alter them such that their deviations from the SM value cancel mutually and yield a
SM-like inclusive cross section.\par
This cancellation of BSM effects in gluon fusion is not merely of academic interest but actually
happens in realistic scenarios. The prime example for this are composite Higgs models.  It was shown
in Refs.\cite{HiggsCouplingsInMCHM} that the effects of a modified Yukawa coupling cancel the
contributions from the top partner loops in a large range of realistic models and make the inclusive
cross section completely insensitive to the top partner mass spectrum. Only a small rescaling of the
cross section is obtained.\par
In the MSSM the cancellation is not as generic as in the composite Higgs models but can happen as
well for large values of the trilinear coupling $A_t$ when it is comparable to the mass of the
second stop. Breaking the degeneracy could even be used to access stealth stops.\par
The main idea of the analysis proposed in the next two sections is to obtain a different relation
between $\kappa_t$ and $\kappa_g$ by making the inclusion of top mass effects necessary. This is
achieved by introducing a new scale to the process that lies above the top mass but below the
potential mass of the top partners. To introduce this scale we demand that the Higgs is produced in
association with a hard jet against which it recoils\footnote{See Refs.\cite{OtherBoostedHiggs} for
  other studies considering the boosted Higgs production in association with a jet to access the
  Higgs couplings}.

\section{Analysis of Higgs + jet}
\label{sec:analysis-p-p}

The amplitude for $p p \to h + \tn{jet}$ is given by
$\mathcal{M}(\kappa_t,\kappa_g)=\kappa_t \mathcal{M}_{IR} + \kappa_g \mathcal{M}_{UV}$, where
$\mathcal{M}_{IR}$ is the amplitude for the top loop contribution given in Refs.\cite{HiggsJet}, and
$\mathcal{M}_{UV}$ is the amplitude stemming from the effective gluon-Higgs interaction. In analogy
to the expression for the inclusive cross section we write
\begin{equation}
  \label{eq:5}
  \frac{\sigma_{p_T^\tn{min}}(\kappa_t,
    \kappa_g)}{\sigma_{p_T^\tn{min}}^{SM}}=\left(\kappa_t+\kappa_g\right)^2 + \delta \kappa_t
  \kappa_g + \epsilon \kappa_g^2\,,
\end{equation}
where $\sigma_{p_T^\tn{min}}$ stands for the cross section for $p p \to h + \tn{jet}$ with a minimal
transverse momentum of the Higgs of $p_T^\tn{min}$. The newly introduced coefficients $\delta$ and
$\epsilon$ quantify the deviation from the inclusive cross section and are calculated using the MSTW
2008 LO PDFs\cite{Martin:2009iq} and the transverse mass $m_T=\sqrt{m_h^2+p_T^2}$ as factorization
and renormalization scale. For $p_T^\tn{min}\to 0$ the process approaches the inclusive production
and the coefficients vanish. However for $p_T^\tn{min}= 800\,\tn{GeV}$ they become
$\delta (\epsilon)\approx 4 (8)$. Of course these large coefficients come with the price of a small
cross section due to the much smaller phase space. As a good compromise between large enough
coefficients $\delta$ and $\epsilon$ and a not too small cross section we found
$p_T^\tn{min}=650\,\tn{GeV}$ by a rough optimization procedure. In order to cancel systematic
uncertainties we divide the boosted cross section by the almost unboosted cross section with
$p_T^\tn{min}=150\,\tn{GeV}$ and take as observable
\begin{equation}
  \label{eq:4}
  \mathcal{R}^0=\frac{\sigma_{650\,\tn{GeV}}(\kappa_t,\kappa_g)\, K_{650\,\tn{GeV}}}{\sigma_{150\,\tn{GeV}}(\kappa_t,\kappa_g) \,K_{150\,\tn{GeV}}}\,,
\end{equation}
where the cross sections are multiplied with the corresponding NLO K-factor obtained from
MCFM-6.6\cite{Campbell:MCFM} to take higher order effects into account.
\par
\begin{figure}[tb]
  \centering
  \includegraphics[width=0.4\textwidth]{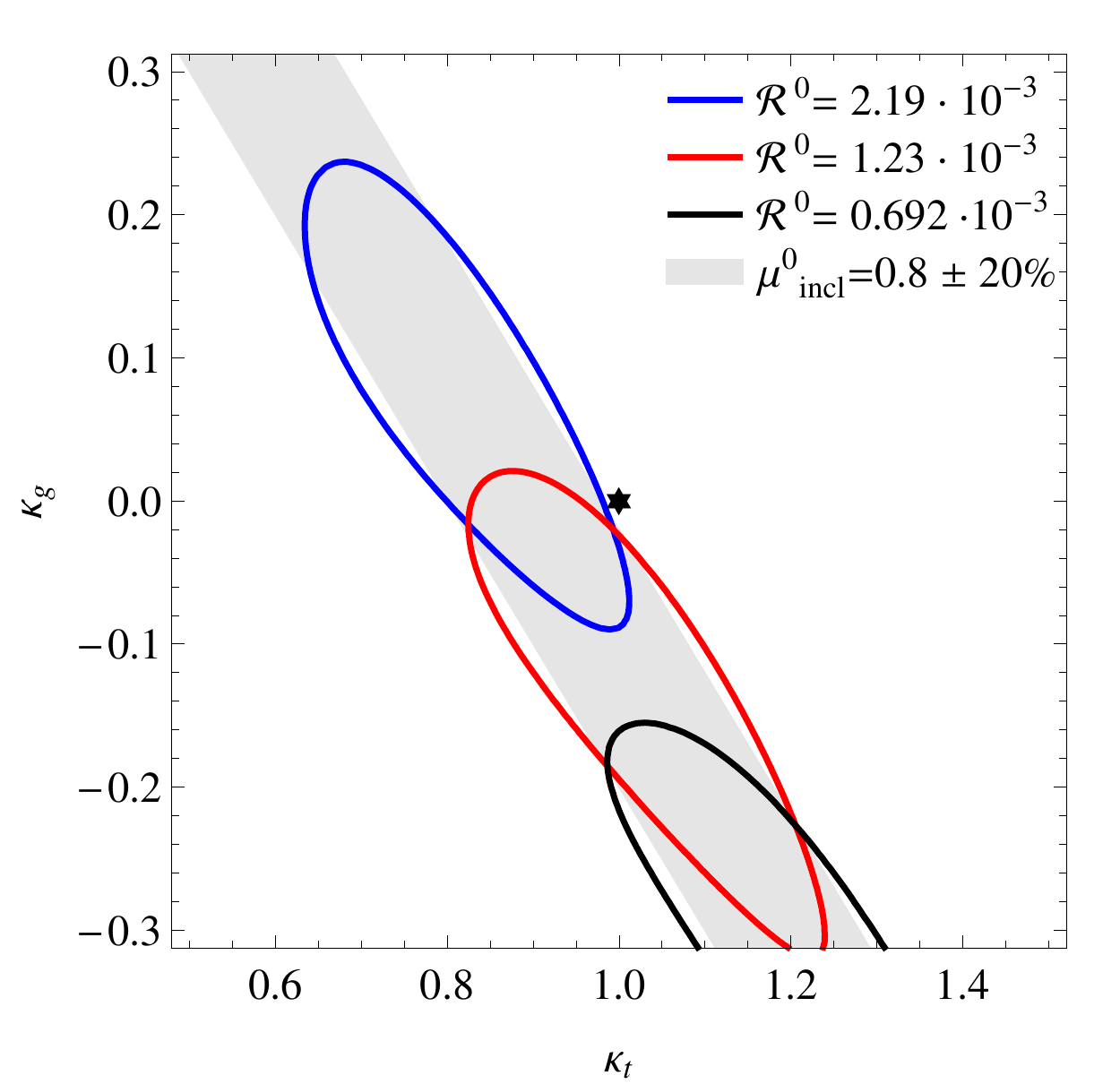}
  \includegraphics[width=0.4\textwidth]{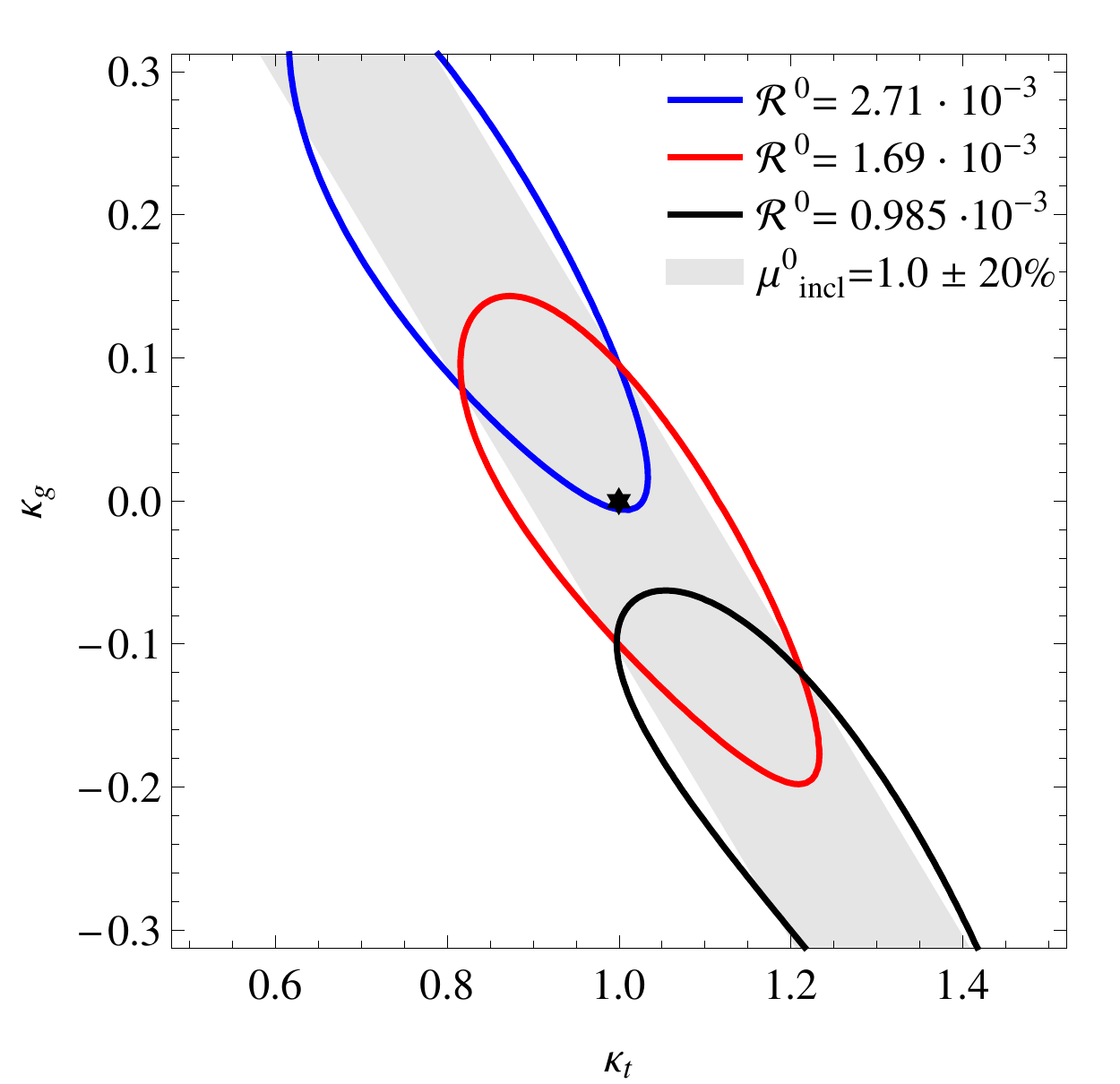}
  \caption{95\% CL contours in the $\kappa_t$-$\kappa_g$ plane assuming an inclusive signal strength
    $\mu^0_\tn{incl}$ of 0.8 (left) and 1.0 (right). The gray band shows the constraint from
    considering only the inclusive cross section and the ellipses the constraint from the $\chi^2$
    fit. The blue, red, and black contour correspond to $\kappa_t=0.8$, 1.0, and 1.2,
    respectively. The corresponding values for $\mathcal{R}^0$ are displayed and the star indicates
    the SM value.}
  \label{fig:ellipses}
\end{figure}

The allowed region in the $\kappa_t$-$\kappa_g$-parameter plane is constrained by performing a
simple $\chi^2$-fit using the inclusive and the boosted cross section as input. For the $95\%$ CL
contours in Fig.~\ref{fig:ellipses} we assumed the center of mass energy $\sqrt{s}=14\,\tn{TeV}$,
integrated luminosity $\mathcal{L}=3\,\tn{ab}^{-1}$, and a systematic uncertainty of 20\% on both
cross sections as well as a statistic error on the boosted cross section. As Higgs decay channel we
chose the decay into $\tau^+\tau^-$ with a SM branching ratio and the reconstruction efficiencies
reported in Ref.\cite{Katz:2010iq}. For more details and references see Ref.\cite{Grojean:2013nya}.

\section{Collider study}
\label{sec:collider-study}

In order to confirm the promising results from the previous section we performed a realistic
collider study. As final state for the signal process we consider $h\to 2\ell + \slashed{E}_T$ with
its most dominant contribution coming from $h\to W_\ell W_\ell^*$ and
$h\to\tau^+_\ell\tau^-_\ell$. As background processes we consider $W$, $Z$, and
$t\bar t + \tn{jets}$ production, where the $W$ bosons (including those of the $t$ decay) are
decaying leptonically and the $Z$ bosons may decay only into $\tau^+\tau^-$ since a $Z$ decaying
into $e$ or $\mu$ can be reconstructed and easily rejected.\par
First the basic event structure---boosted Higgs and recoiling jet---is demanded by reconstructing
the Higgs transverse momentum $p_T^h=p_T^{\ell_1}+p_T^{\ell_2}+\slashed{E}_T$ and rejecting events
with $p_T^h<200\,\tn{GeV}$. Moreover a fat jet with $p_T>200\,\tn{GeV}$ is required. Next we observe
that a spin correlation in the $h\to W_\ell W^*_\ell$ channel leads to $\slashed{E}_T$ lying outside
the cone defined by the two leptons. In the $h\to \tau^+\tau^-$ channel no such correlation exists
and $\slashed{E}_T$ lies mostly inside this cone. This criterium allows us to distinguish the
channels and tailor the analysis accordingly.\par
In the $h\to W_\ell W^*_\ell$ channel we calculate
$m_{T,\ell\ell}^2=m_{\ell\ell}^2+2(E_{T,\ell\ell} \slashed{E}_T-\boldsymbol{p}_{T,\ell\ell}\cdot
\slashed{\boldsymbol{p}}_T)$
which gives a lower bound on the Higgs mass and reject all events with $m_{T,\ell\ell} > m_h$. In
addition we demand that the two leptons are close by: $\Delta R_{\ell\ell}\le 0.4$. In the end we
achieve $S/B\sim 0.4$ and $S/\sqrt{B}>6$ for $\mathcal{L}=300\,\tn{fb}^{-1}$ in this channel.\par
In the $h\to \tau^+\tau^-$ channel a large fraction of the $t\bar t$ and $W$ background can be
rejected by vetoing events with a dilepton mass $m_{\ell\ell}>70\,\tn{GeV}$. Eventually the Higgs
mass can be reconstructed by the collinear approximation which assumes that the neutrino momenta are
parallel to the reconstructed leptons and make up all missing energy. Requiring the reconstructed
mass to lie within 10\,GeV of the actual Higgs mass yields $S/B\sim 0.4$ and $S/\sqrt{B}>9$ for
$\mathcal{L}=300\,\tn{fb}^{-1}$.\par
\begin{figure}[tb]
  \centering
  \begin{tikzpicture}
    \node[right] at (0,43) {\includegraphics[height=120pt, trim=0pt 0pt 60pt 0pt, clip]{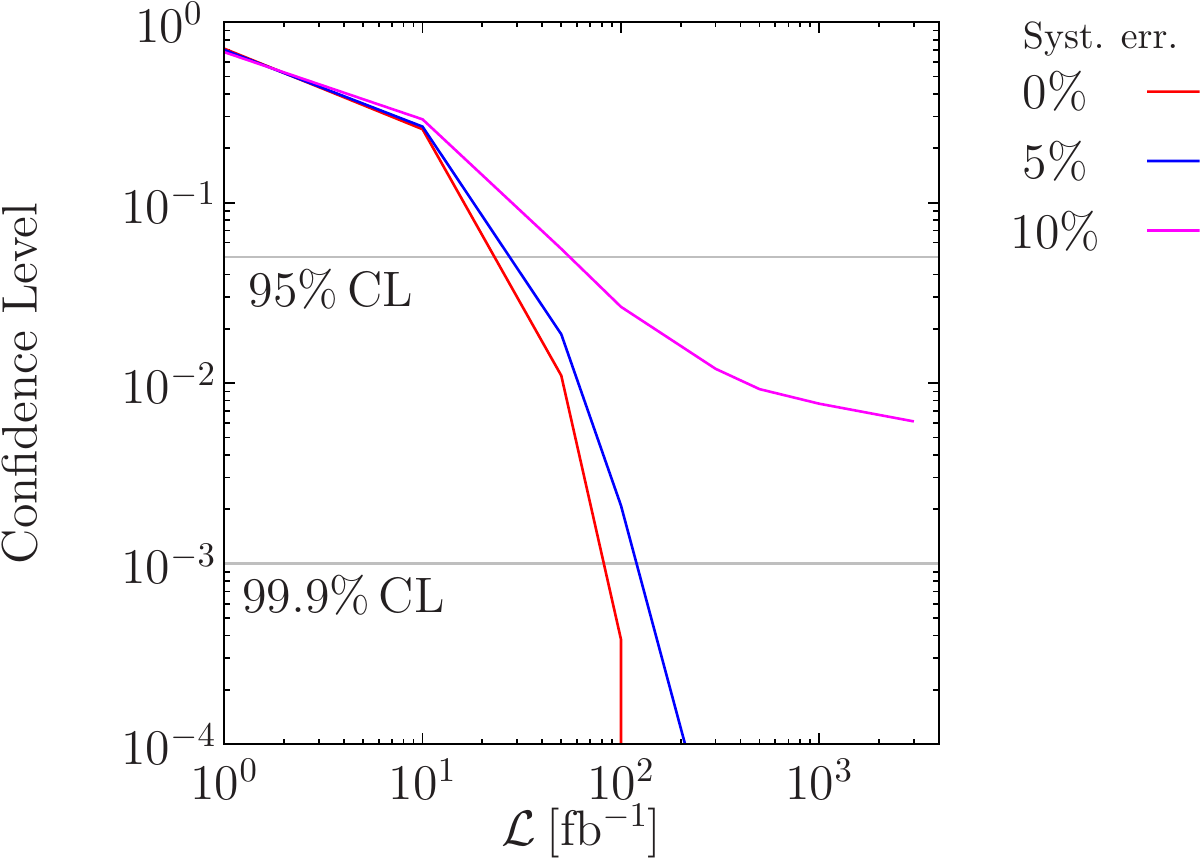}};
    \node at (0.46\textwidth,43) {\includegraphics[height=120pt, trim=0pt 0pt 60pt 0pt, clip]{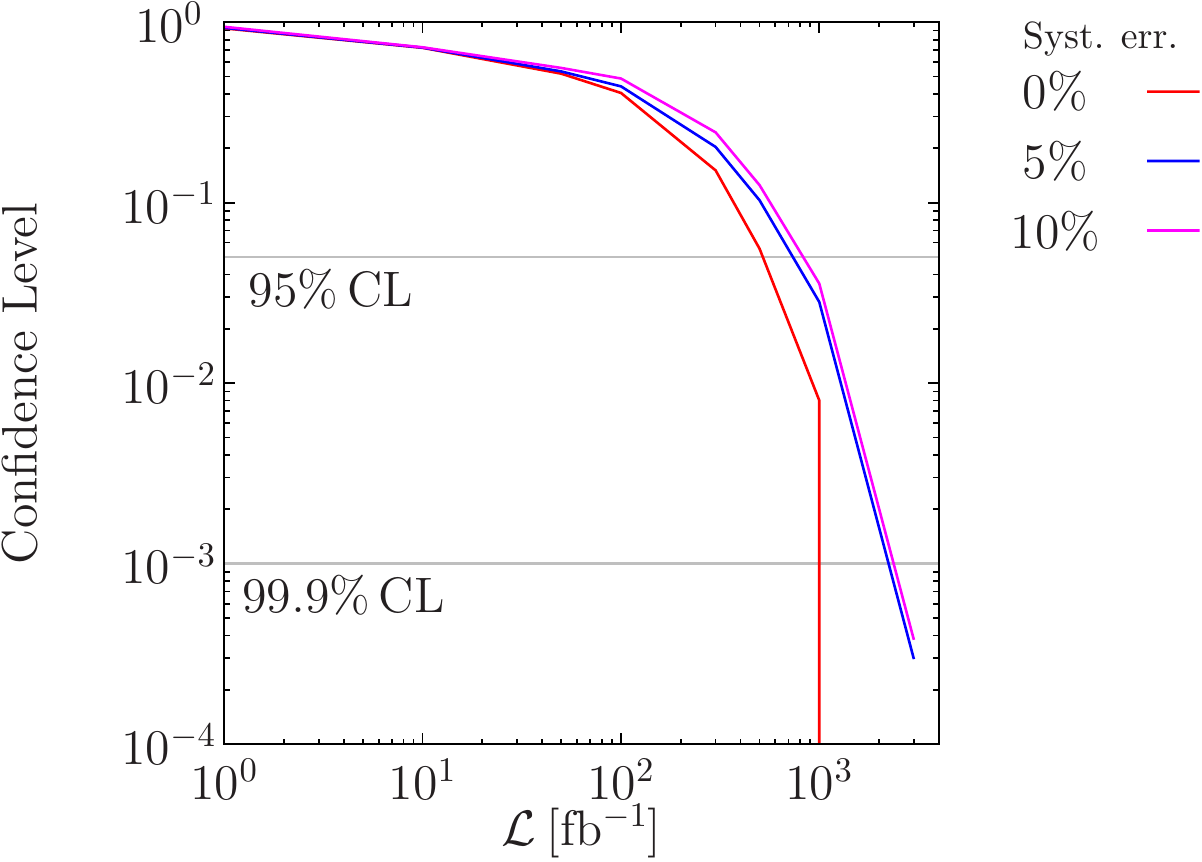}};
    \node[inner sep=0] at (0.77\textwidth,46) {\includegraphics[height=135pt, trim=0pt 40pt 0pt 0pt, clip]{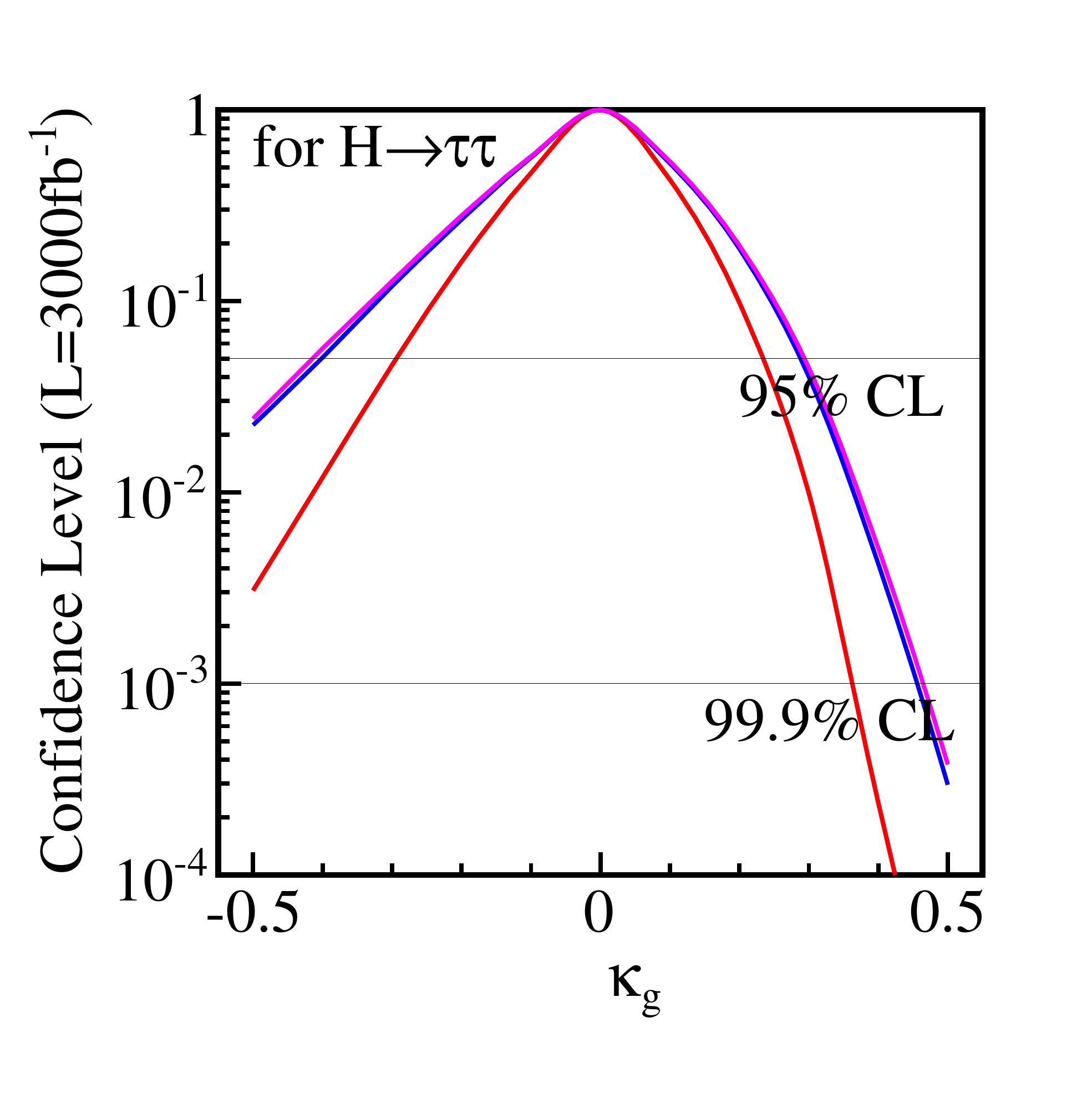}};
    \node[inner sep=0, left] at (\textwidth, 63) {\includegraphics[height=60pt, trim=290pt 170pt 0pt 0pt, clip]{pHT_tt_10_bkg_CLs}};
  \end{tikzpicture}
  \caption{p-values for boosted Higgs with $\kappa_t=1.0$ against background processes (left) and
    BSM signal with $\kappa_t=0.5$ against SM signal (center) as function of luminosity. The right
    panel shows the p-values as function of $\kappa_t$ for an integrated luminosity of
    $3\,\tn{ab}^{-1}$. For these plots only the $h\to\tau_\ell\tau_\ell$ channel was considered.}
  \label{fig:CLsfig}
\end{figure}

For the $h\to \tau^+\tau^-$ channel we performed a binned likelihood ratio fit using the $CL_s$
method\cite{Junk:1999kv} under the assumption of the worst-case scenario without deviations in the
inclusive cross section measurement. The results are shown in Fig.~\ref{fig:CLsfig}. In this
scenario the boosted Higgs can be seen against the background at 95\% CL with an integrated
luminosity of less than $100\,\tn{fb}^{-1}$. A more detailed description of the analysis and further
references can be found in Ref.\cite{Schlaffer:2014osa}.

\section{Conclusion}
\label{sec:conclusion}

We used boosted Higgs production in gluon fusion to disentangle the contributions of a modified top
Yukawa coupling and of new top partners quantified by $\kappa_t$ and $\kappa_g$, respectively. By
combining the inclusive and the boosted cross section which have a different dependence on
$\kappa_t$ and $\kappa_g$ the allowed region in the $\kappa_t$-$\kappa_g$-plane can be
constrained. Assuming the worst case scenario with a SM inclusive cross section and a systematic
uncertainty of 10\%, $\kappa_g$ can be constrained at 95\% CL to $-0.4\le\kappa_g\le 0.3$ by
considering only the decay $h\to\tau^+_\ell\tau^-_\ell$ with an integrated luminosity of
$3\,\tn{ab}^{-1}$. Therefore the boosted Higgs channel is an interesting alternative to determine
the top Yukawa coupling independently of the $t\bar t h$ channel.

\section*{Acknowledgments}

I would like to thank the organizers of the 27th Rencontres de Blois for the interesting workshop. I
am grateful to Christophe Grojean, Ennio Salvioni, Michael Spannowsky, Michihisa Takeuchi, Andreas
Weiler, and Chris Wymant for their contributions to the projects this talk is based on. Furthermore
I acknowledge the funding by the Joachim-Herz-Stiftung.


\section*{References}

\begin{thebibliography}{99}

\bibitem{HLET}
  J.~R.~Ellis, M.~K.~Gaillard and D.~V.~Nanopoulos,
  Nucl.\ Phys.\ B {\bf 106} (1976) 292;
%
  M.~A.~Shifman, A.~I.~Vainshtein, M.~B.~Voloshin and V.~I.~Zakharov,
  Sov.\ J.\ Nucl.\ Phys.\  {\bf 30} (1979) 711
   [Yad.\ Fiz.\  {\bf 30} (1979) 1368].

\bibitem{Gillioz:2012se}
  M.~Gillioz, R.~Gr\"ober, C.~Grojean, M.~M\"uhlleitner and E.~Salvioni,
  JHEP {\bf 1210} (2012) 004,
  \hhref{1206.7120} [hep-ph].

\bibitem{OtherBoostedHiggs}
  R.~V.~Harlander and T.~Neumann,
  Phys.\ Rev.\ D {\bf 88} (2013) 074015,
  \hhref{1308.2225} [hep-ph].
%
  A.~Banfi, A.~Martin and V.~Sanz,
  \hhref{1308.4771} [hep-ph].
%
  A.~Azatov and A.~Paul,
  JHEP {\bf 1401} (2014) 014,
  \hhref{1309.5273} [hep-ph].


\bibitem{HiggsCouplingsInMCHM}
  A.~Falkowski,
  Phys.\ Rev.\ D {\bf 77} (2008) 055018,
  \hhref{0711.0828} [hep-ph].
%
  I.~Low and A.~Vichi,
  Phys.\ Rev.\ D {\bf 84} (2011) 045019,
  \hhref{1010.2753} [hep-ph].
%
  A.~Azatov and J.~Galloway,
  Phys.\ Rev.\ D {\bf 85} (2012) 055013,
  \hhref{1110.5646} [hep-ph].
%
  C.~Delaunay, C.~Grojean and G.~Perez,
  JHEP {\bf 1309} (2013) 090,
  \hhref{1303.5701} [hep-ph].
%
  M.~Montull, F.~Riva, E.~Salvioni and R.~Torre,
  Phys.\ Rev.\ D {\bf 88} (2013) 095006,
  \hhref{1308.0559} [hep-ph].

\bibitem{HiggsJet}
  R.~K.~Ellis, I.~Hinchliffe, M.~Soldate and J.~J.~van der Bij,
  Nucl.\ Phys.\ B {\bf 297} (1988) 221.
%
  U.~Baur and E.~W.~N.~Glover,
  Nucl.\ Phys.\ B {\bf 339} (1990) 38.

\bibitem{Martin:2009iq}
  A.~D.~Martin, W.~J.~Stirling, R.~S.~Thorne and G.~Watt,
  Eur.\ Phys.\ J.\ C {\bf 63} (2009) 189,
  \hhref{0901.0002} [hep-ph].

\bibitem{Campbell:MCFM}
  J.~M.~Campbell, R.~K.~Ellis and C.~Williams,
  MCFM web page \url{http://mcfm.fnal.gov/}.

\bibitem{Katz:2010iq}
  A.~Katz, M.~Son and B.~Tweedie,
  Phys.\ Rev.\ D {\bf 83} (2011) 114033,
  \hhref{1011.4523} [hep-ph].

\bibitem{Grojean:2013nya}
  C.~Grojean, E.~Salvioni, M.~Schlaffer and A.~Weiler,
  JHEP {\bf 1405} (2014) 022
  \hhref{1312.3317} [hep-ph].




\bibitem{Junk:1999kv}
  T.~Junk,
  Nucl.\ Instrum.\ Meth.\ A {\bf 434} (1999) 435
  [hep-ex/9902006].

\bibitem{Schlaffer:2014osa}
  M.~Schlaffer, M.~Spannowsky, M.~Takeuchi, A.~Weiler and C.~Wymant,
  Eur.\ Phys.\ J.\ C {\bf 74} (2014) 10,  3120
  \hhref{1405.4295} [hep-ph].




\end{thebibliography}
\end{document}